\documentclass[superscriptaddress,amssymb,amsmath,nobibnotes,aps,prd,eqsecnum,twocolumn,showpacs,
nofootinbib]{revtex4}
\usepackage{graphics,color}
\usepackage{bm}
\usepackage{graphicx}
\usepackage{amsmath}
\usepackage{amssymb}
\usepackage{enumitem}
\usepackage{epstopdf}
\usepackage[colorlinks=true]{hyperref}
\allowdisplaybreaks	

\def\be{\begin{equation}}
\def\ee{\end{equation}}
\def\bea{\begin{eqnarray}}
\def\eea{\end{eqnarray}}
\def\l{\left}
\def\r{\right}
\def\f{\frac}

\def\nn{\nonumber \\}

\begin{document}

\title{Crossing SNe Ia and BAO observational constraints with local ones in hybrid metric-Palatini gravity}
\author{Iker Leanizbarrutia}
\email{iker.leanizbarrutia@ehu.es}
\affiliation{Fisika Teorikoaren eta Zientziaren Historia Saila, Zientzia eta Teknologia Fakultatea,\\
Euskal Herriko Unibertsitatea, 644 Posta Kutxatila, 48080 Bilbao, Spain}
\author{Francisco S.N. Lobo}
\email{fslobo@fc.ul.pt}
\affiliation{Instituto de Astrof\'{\i}sica e Ci\^{e}ncias do Espa\c{c}o, Faculdade de
Ci\^encias da Universidade de Lisboa, Edif\'{\i}cio C8, Campo Grande,
P-1749-016 Lisbon, Portugal}
\author{Diego S\'aez-G\'omez}
\email{saez@ice.csic.es}
\affiliation{Institut de Ci\`{e}ncies de l'Espai, ICE/CSIC-IEEC, Campus UAB, Carrer de Can Magrans s/n, 08193 Bellaterra (Barcelona), Spain}

\date{\today}

%%%%%%%%%%%%%%%%%%%%%%%%%%%%%%%%%%%%%%%%%%%%%%%%%%%%%%%%%%%%%%%%%%%%%%%%%%%%%%%%%%%
\begin{abstract}
In this paper, we consider the recently proposed hybrid metric-Palatini gravitational theory, which consists of adding to the Einstein-Hilbert Lagrangian an $f(\mathcal{R})$ term constructed \`{a} la Palatini. Using the respective dynamically equivalent scalar-tensor representation, we explore the cosmological evolution of a specific model, given by $f(\mathcal{R}) \propto \mathcal{R}^2$, and obtain constraints on the free parameters by using different sources of cosmological data. The viability of the model is analysed by combining the conditions imposed by the Supernovae Ia and Baryonic Acoustic Oscillations data and the results are compared with the local constraints.
\end{abstract}
%%%%%%%%%%%%%%%%%%%%%%%%%%%%%%%%%%%%%%%%%%%%%%%%%%%%%%%%%%%%%%%%%%%%%%%%%%%%%%%%%%%

\pacs{04.50.Kd, 98.80.-k, 95.36.+x}

\maketitle
%\tableofcontents

%%%%%%%%%%%%%%%%%%%%%%%%%%%%%%%%%%
\section{Introduction}
%%%%%%%%%%%%%%%%%%%%%%%%%%%%%%%%%%

The discovery of the late-time accelerated expansion of the Universe \cite{Perlmutter:1998np,Riess:1998cb} has motivated a tremendous amount of research on dark energy models \cite{Copeland:2006wr} and modified theories of gravity \cite{Sotiriou:2008rp,DeFelice:2010aj,Clifton:2011jh,Nojiri:2010wj,bookFR,reviewAandD} as a possible source for this cosmic speed-up. In the latter framework, it is assumed that at large scales Einstein's General Relativity (GR) breaks down, and one needs to introduce new degrees of freedom to the gravitational sector. In this context, $f(R)$ gravity has been widely explored to address this problem, where essentially two approaches have been analysed in the literature. The first approach consists on varying the action with respect to the metric \cite{Sotiriou:2008rp}, and the second approach, denoted by the Palatini formalism, consists on  treating the metric and the connection as separate variables \cite{Olmo:2011uz}, and these are used in varying the action to obtain the respective field equations. In $f(R)$ gravity, both formalisms are related to two different theories, which easily be verified by the respective scalar-tensor representations. In fact, one can show that the $f(R)$ metric formalism corresponds to a Brans-Dicke theory (in the presence of a potential) with a Brans-Dicke parameter given by $w_{BD}=0$ and the Palatini formalism to a Brans-Dicke parameter with $w_{BD}=-3/2$. Moreover, within the metric approach, specific viable models have been proposed, which are capable of keeping the GR results at local scales \cite{Hu:2007nk}  and provide good fits when compared with cosmological data \cite{delaCruz-Dombriz:2015tye}.

Now, recently, a novel approach to $f(R)$ gravity has been proposed that consists of adding to the metric Einstein-Hilbert Lagrangian an $f(\cal{R})$ term constructed \`{a} la Palatini \cite{Harko:2011nh}. Using the respective dynamically equivalent scalar-tensor representation, it was shown that the theory can pass the Solar System observational constraints even if the scalar field is very light. This implies the existence of a long-range scalar field, which is able to modify the cosmological and galactic dynamics, but leaves the Solar System unaffected. This has motivated further research in this promising model.
In fact, it was shown that the theory can also be formulated in terms of the quantity $X\equiv \kappa^2 T+R$, where $T$ and $R$ are the traces of the energy-momentum and Ricci tensors, respectively, and the variable $X$ represents the deviation with respect to the field equation trace of GR. The cosmological applications of this hybrid metric-Palatini gravitational theory were explored \cite{Capozziello:2012ny}, and criteria to obtain cosmic acceleration were discussed. Several classes of dynamical cosmological solutions, depending on the functional form 
of the effective scalar field potential, describing both accelerating and 
decelerating Universes were explicitly obtained and a dynamical system was explored, which was further explored in \cite{Carloni:2015bua}. The problem of dark matter was also addressed in the context of the hybrid theory \cite{Capozziello:2012qt,Capozziello:2013yha,Capozziello:2013uya}. We refer the reader to Ref. \cite{Capozziello:2015lza} for a recent review.

The cosmological perturbation equations were derived and applied to uncover the nature of the propagating scalar degree of freedom and the signatures these models predict in the large-scale structure.  
The evolution of the linear perturbations in the hybrid theory was analysed in \cite{Lima:2014aza}, where the full set of linearized evolution equations, for the perturbed potentials in the Newtonian and synchronous gauges, were derived in the Jordan frame. It was concluded, that for the specific model used, that the main deviations from GR arise in the distant past, with an oscillatory signature in the ratio between the Newtonian potentials.
Furthermore, two particular models of the hybrid metric-Palatini theory were introduced \cite{Lima:2015nma}, and their background evolution was explored. It was shown explicitly that one recovers GR with an effective Cosmological Constant at late times. This is due to the fact that the Palatini Ricci scalar evolves towards and asymptotically settles at the minimum of its effective potential during cosmological evolution.
  
In this work, we consider the respective dynamically equivalent scalar-tensor representation of the hybrid metric-Palatini theory and explore the cosmological evolution of a specific model, given by $f(\mathcal{R}) \propto \mathcal{R}^2$. Furthermore, we obtain constraints on the free parameters by using different sources of cosmological data. More specifically, the viability of the model is analysed by combining the conditions imposed by Supernovae Ia and Baryonic Acoustic Oscillations (BAO) data and the results are compared with the local constraints.
  
This paper is outlined in the following manner: In Section \ref{sectionII}, we present the general formalism of the hybrid metric-Palatini theory, in particular, the action and the field equations in the curvature approach and the scalar-tensor representation. In Section \ref{sectionIII}, we consider the weak field limit of the hybrid gravity, and present the condition to pass the local tests. In Section \ref{sectionIV}, we consider a particular model which is then confronted with cosmological observations. More specifically, the cosmological evolution of the model is analysed and constraints on the free parameters are obtained by using different sources of data, and the viability of the model is analysed by combining the cosmological and local constraints. We finally present and discuss our results in Section \ref{sectionV} and conclude in Section \ref{sectionConclusion}.

%%%%%%%%%%%%%%%%%%%%%%%%%%%%%%%%%%%%%%%%%%%%%%%%%%%%%%%%%%%
\section{Hybrid metric-Palatini gravity}\label{sectionII}
%%%%%%%%%%%%%%%%%%%%%%%%%%%%%%%%%%%%%%%%%%%%%%%%%%%%%%%%%%%

%%%%%%%%%%%%%%%%%%%%%%%%%%%%%%%%%%%%%%%%%%%%%%%%%%%%%%%%%%%
\subsection{Action and field equations}
%%%%%%%%%%%%%%%%%%%%%%%%%%%%%%%%%%%%%%%%%%%%%%%%%%%%%%%%%%%

In the present manuscript, we consider a class of hybrid metric-Palatini gravity given by the Hilbert-Einstein term and an arbitrary function of the curvature scalar, which is constructed \`{a} la Palatini. This can be expressed by the following Lagrangian \cite{Harko:2011nh}:
\be
\mathcal{S}=\frac{1}{2\kappa^2}\int{}d^4x\sqrt{-g}\l[R+f(\mathcal{R})\r]+\int{}d^4x\sqrt{-g}\mathcal{L}_m(g_{\mu\nu},\psi_i),
\label{1.1}
\ee
where $\kappa^2=8\pi G$, $R$ is the Ricci scalar defined in terms of the Levi-Civita connection of the metric $g_{\mu\nu}$ while  $\mathcal{R}\equiv g^{\mu\nu}\mathcal{R}_{\mu\nu}$ is the Palatini curvature, where the Ricci tensor is defined in terms of an independent connection $\tilde{\Gamma}^\alpha_{\mu\nu}$ as
\be
\mathcal{R}_{\mu\nu}=\tilde{\Gamma}^{\lambda}_{\mu\nu,\lambda}-\tilde{\Gamma}^{\lambda}_{\mu\lambda,\nu}+\tilde{\Gamma}^{\lambda}_{\lambda\sigma}\tilde{\Gamma}^{\sigma}_{\mu\nu}-\tilde{\Gamma}^{\lambda}_{\mu\sigma}\tilde{\Gamma}^{\sigma}_{\lambda\nu}\ .
\label{1.2}
\ee

By varying the action (\ref{1.1}) with respect to the metric, the following field equation is obtained:
\be
R_{\mu\nu}-\frac{1}{2}g_{\mu\nu}R+f_{\mathcal{R}}\mathcal{R}_{\mu\nu}-\frac{1}{2}f(\mathcal{R})g_{\mu\nu}=\kappa^2T_{\mu\nu}\ ,
\label{1.3}
\ee
where $f_{\mathcal{R}}=\partial f(\mathcal{R})/\partial \mathcal{R}$ and the energy-momentum tensor is defined as usual by $T_{\mu\nu}=\frac{-2}{\sqrt{-g}}\frac{\delta (\sqrt{-g}\mathcal{L}_m)}{\delta g^{\mu\nu}}$. Then, varying the action (\ref{1.1}) with respect to the independent connection $\tilde{\Gamma}^{\lambda}_{\mu\nu}$ yields
\be
\tilde{\nabla}_{\lambda}\l[\sqrt{-g}\l(\delta^{\lambda}_{\alpha}f_{\mathcal{R}}g^{\beta\gamma}-\frac{1}{2}\delta^{\beta}_{\alpha}f_{\mathcal{R}}g^{\lambda\gamma}-\frac{1}{2}\delta^{\gamma}_{\alpha}f_{\mathcal{R}}g^{\beta\lambda}\r)\r]=0\,.
\label{1.4}
\ee

It is straightforward to show that the connection $\tilde{\Gamma}^{\lambda}_{\mu\nu}$ is compatible with the metric $f_{\mathcal{R}}g_{\mu\nu}$, which is basically a conformal transformation of the metric $g_{\mu\nu}$. Hence, both Ricci tensors are related by a conformal factor, such that:
\be
\mathcal{R}_{\mu\nu}=R_{\mu\nu}+\frac{3}{2}\frac{f_{\mathcal{R},\mu}f_{\mathcal{R},\nu}}{f_{\mathcal{R}}^2}-\frac{f_{\mathcal{R};\mu\nu}}{f_{\mathcal{R}}}-\frac{1}{2}\frac{g_{\mu\nu}\Box f_{\mathcal{R}}}{f_{\mathcal{R}}}\ .
\label{1.5}
\ee

In addition, by taking the trace of the field equations (\ref{1.3}), the curvature $\mathcal{R}$ can be expressed in terms of the curvature $R$ and the trace of the energy-momentum tensor \cite{Capozziello:2012ny}:
\be
f_{\mathcal{R}}\mathcal{R}-2f(\mathcal{R})=\kappa^2T+R=X\ ,
\label{1.6}
\ee
which can be solved algebraically to obtain $\mathcal{R}=\mathcal{R}(X)$. Note that the trace of the Einstein equations gives $\kappa^2T+R=0$, such that the new variable $X$ may be used to measure deviations from GR. Moreover, the field equations can be expressed in terms of the metric $g_{\mu\nu}$ and $X$, such that for a particular cosmological solution the corresponding action can be easily reconstructed by inverting the equation (\ref{1.6}). 

In the next section, the scalar-tensor representation of the theory is explored, which yields interesting properties that are shown in the subsequent sections and allows us to study the cosmological behaviour of a viable model by using such a framework.

%%%%%%%%%%%%%%%%%%%%%%%%%%%%%%%%%%%%%%%%%%%%%%%%%%%%%%%%%%%
\subsection{Scalar-tensor representation}
%%%%%%%%%%%%%%%%%%%%%%%%%%%%%%%%%%%%%%%%%%%%%%%%%%%%%%%%%%%

As in the case of the metric and Palatini $f(R)$ gravities, the action (\ref{1.1}) can be expressed in terms of a scalar field, which simplifies the analysis, such as the study of the Newtonian law corrections or the study of cosmological solutions within hybrid $f(\mathcal{R})$ gravity, as shown in the next sections. In such a case, the action is given by \cite{Harko:2011nh}
\bea
\mathcal{S}=\frac{1}{2\kappa^2}\int{}d^4x\sqrt{-g}\l[R+\phi\mathcal{R}-V(\phi)\r]
   \nonumber \\
+\int{}d^4x\sqrt{-g}\mathcal{L}_m(g_{\mu\nu},\psi_i)\ ,
\label{1.7}
\eea
Then, by varying the action with respect to the scalar field $\phi$, it yields,
\be
\mathcal{R}-\frac{\partial V}{\partial\phi}=0\quad \Rightarrow \quad \phi=\phi(\mathcal{R})\ .
\label{1.8}
\ee
Hence, the original action (\ref{1.1}) is recovered provided that Eq.~(\ref{1.8}) is invertible, 
\be
f(\mathcal{R})=\phi(\mathcal{R})\mathcal{R}-V\l(\phi(\mathcal{R})\r)\ ,
\label{1.9a}
\ee
where the scalar field and the potential are given by
\be
\phi=f_{\mathcal{R}}\ , \qquad V(\phi)=\mathcal{R}f_{\mathcal{R}}-f(\mathcal{R})\,,
\label{1.9}
\ee
respectively.

Besides Eq.~(\ref{1.8}), the field equations are obtained by varying the action (\ref{1.7}) with respect to the metric and the independent connection, leading to:
\be
R_{\mu\nu}-\frac{1}{2}g_{\mu\nu}R+\phi\mathcal{R}_{\mu\nu}-\frac{1}{2}\l[\phi g_{\mu\nu}-V(\phi)\r]=\kappa^2T_{\mu\nu}\ , \label{1.10a}
\ee
and
\be
\tilde{\nabla}\l(\sqrt{-g}\phi g^{\mu\nu}\r)=0 \,,
\label{1.10b}
\ee
respectively.
These equations are equivalent to those given in Eqs. (\ref{1.3}) and (\ref{1.4}) as can easily be shown by using Eq. (\ref{1.9}). As stated above, the solution for Eq. (\ref{1.10b}) is given by the Levi-Civita connection of the metric $\phi g_{\mu\nu}$, that is conformally related to $g_{\mu\nu}$, such that the Ricci tensor $R_{\mu\nu}$ is conformally transformed (\ref{1.5}), and we have
\be
\mathcal{R}_{\mu\nu}=R_{\mu\nu}+\frac{3}{2\phi^2}\partial_{\mu}\phi\partial_{\nu}\phi-\frac{1}{\phi}\l(\nabla_{\mu}\nabla_{\nu}\phi+\frac{1}{2}\Box\phi\r)\,,
\ee
\bea
\mathcal{R}=g^{\mu\nu}\mathcal{R}_{\mu\nu}=R+\frac{3}{2\phi^2}\partial_{\mu}\phi\partial^{\mu}\phi-\frac{3}{\phi}\Box\phi\ .
\label{1.11}
\eea
Here the last term in the expression of $\mathcal{R}$ is a total derivative, such that the action (\ref{1.7}) is given by:
\bea
\mathcal{S}&=&\frac{1}{2\kappa^2}\int{}d^4x\sqrt{-g}\l[(1+\phi)R+\frac{3}{2\phi}\partial_{\mu}\phi\partial_{\nu}\phi-V(\phi)\r]
    \nonumber  \\
&&+\int{}d^4x\sqrt{-g}\mathcal{L}_m(g_{\mu\nu},\psi_i)\ .
\label{1.12}
\eea
This is the action of a non-minimally coupled scalar field with a non-canonical kinetic term which mimics somehow the Brans-Dicke theory except for the self-interacting term of the scalar field and the coupling to the curvature. The field equations become
\bea
 R_{\mu\nu}&-&\frac{1}{2}g_{\mu\nu}R=\frac{1}{(1+\phi)}\Big[\kappa^2T_{\mu\nu}+\nabla_{\mu}\nabla_{\nu}\phi-g_{\mu\nu}\Box\phi  \label{1.13a}
    \nonumber  \\ 
 &-&\frac{3}{2\phi}\partial_{\mu}\phi\partial_{\nu}\phi+\frac{3}{4\phi}g_{\mu\nu}\partial_{\lambda}\phi\partial^{\lambda}\phi-\frac{1}{2}g_{\mu\nu}V\Big],
\eea 
\bea 
 \Box\phi+\frac{1}{2\phi}\partial_{\mu}\phi\partial^{\mu}\phi+\frac{\phi}{3}\l[2V-(1+\phi)V_{\phi}\r]=\frac{\kappa^2}{3}\phi T ,
 \label{1.13}
\eea
respectively. In comparison to the pure Palatini case, the scalar field is dynamical when assuming the hybrid action (\ref{1.1}), as shown by Eq. (\ref{1.13}). 

In the next section, we review the weak field limit of $R+f(\mathcal{R})$ gravity and consider a particular model which is then confronted with cosmological observations.

%%%%%%%%%%%%%%%%%%%%%%%%%%%%%%%%%%%%%%%%%%%%%%%%%%%%%%%%%%%
\section{Newton law corrections in $R+f(\mathcal{R})$ gravity} 
\label{sectionIII}
%%%%%%%%%%%%%%%%%%%%%%%%%%%%%%%%%%%%%%%%%%%%%%%%%%%%%%%%%%%

In order to study the corrections induced by the extra scalar field at local scales such as the Earth or the Solar System, we proceed to analyse the correction to the weak field limit, similarly as performed in Ref.~\cite{Olmo:2005hc} for the metric/Palatini case. Then, one considers the metric to be quasi-Minkowski at local scales where the cosmological evolution has negligible effects and the scalar field is also approximated to its asymptotic value, which is assumed constant around the present time. Hence, the weak field limit may be expressed as follows:
\bea
g_{\mu\nu} &\approx & \eta_{\mu\nu}+h_{\mu\nu}(\bf{x})\ , \quad |h_{\mu\nu}|\ll 1\ , \nn
\phi&\approx& \phi_0+\tilde{\phi}(\bf{x})\ , \quad |\tilde{\phi}|\ll 1\ .
  \nonumber
\label{2.1}
\eea
As shown in \cite{Harko:2011nh} for the hybrid action (\ref{1.1}), the field equations (\ref{1.13a}) and (\ref{1.13}) for $h_{\mu\nu}$ and $\phi_1(\bf{x})$ at linear order yield:
\be
\nabla^2h_{\mu\nu}=-\frac{2}{1+\phi_0}\l( T_{\mu\nu}-\frac{1}{2}\eta_{\mu\nu}T\r)+\frac{V_0+\nabla^2\tilde{\phi}}{2(1+\phi_0)}\eta_{\mu\nu}\,,
\ee
and
\bea
(\nabla^2-m_{\phi}^2)\phi = \frac{\kappa^2\phi_0}{3}\rho_m\ .
\label{2.2}
\eea
Here $\nabla^2$ is the 3D Laplacian of flat space and 
\be
m_{\phi}^2=\frac{1}{3}\left[2V-V_{\phi}-\phi(1+\phi)V_{\phi\phi}\right]\big|_{\phi=\phi_{0}} \,,
\ee
%$m_{\phi}^2=\frac{(2V-V_{\phi}-\phi(1+\phi)V_{\phi\phi})}{3}|_{\phi=\phi_{0}}$ 
is the effective mass of the scalar field. The general solution assuming spherical symmetry and far from the sources was found in Ref.~\cite{Harko:2011nh}, and yields:
\bea
h_{00}(r)&=&\frac{2G_{\rm eff}M}{r}+\frac{V_{0}}{1+\phi_0}\frac{r^2}{6}\ , \\
h_{ij}(r)&=& \l(\frac{2\gamma G_{\rm eff}M}{r}-\frac{V_{0}}{1+\phi_0}\frac{r^2}{6}\r)\ ,
\label{2.3}
\eea
where $M=\int d^3x \rho$ is the mass of the source, $G_{\rm eff}$ is the effective Newtonian constant and $\gamma$ is the post-Newtonian parameter, both given by:
\bea
G_{\rm eff}&=&\frac{G}{1+\phi_0}\l(1-\frac{\phi_0}{3}e^{-m_{\phi}r}\r)\ , \nn
\gamma&=&\frac{1+(\phi_0/3)e^{-m_{\phi}r}}{1-(\phi_0/3)e^{-m_{\phi}r}}\ .
\label{2.4}
\eea
In order to avoid large corrections at scales of the Earth or the Solar System, the effective Newtonian constant has to be approximately $G$ and the post-Newtonian parameter $\gamma\approx 1$. Hence, contrary to the case of metric $f(R)$ gravity where a large mass of the scalar field $m_{\phi}$ is required, which scales with the curvature through the chameleon mechanism in order to satisfy the observational constraints \cite{Khoury:2003rn}, in the hybrid gravity case described by the action (\ref{1.1}), just a small value of $|\phi_0|\ll 1$ is required. Nevertheless, a positive mass $m_{\phi}^2>0$ is also needed in order to avoid instabilities. 

Let us now introduce the model that is constrained in the next section in order to test the local constraints provided by (\ref{2.4}),
\be
f(\mathcal{R})= \frac{\mathcal{R}^2}{4V_1}-V_0\ ,
\label{2.5}
\ee
where $\{V_0, V_1\}$ are constants. By using the scalar-tensor representation (\ref{1.7}) or equivalently (\ref{1.12}), the corresponding self-interacting term for the scalar field is given by
\be
V(\phi)=V_0+V_1\phi^2\ .
\label{2.6}
\ee
Hence, in order to avoid deviations from Newton's law at local scales (\ref{2.4}), one requires to have $\phi_0\ll 1$ while the scalar mass should be positive:
\be
m_{\phi}^2=\frac{2}{3}(V_0-2V_1\phi)>0, \quad  \Rightarrow \quad \phi<V_0/2V_1 \,.
\label{2.7}
\ee
Thus, the model (\ref{2.5}) is consistent at local scales as far as the above conditions hold. In the next section, the cosmological evolution of the model (\ref{2.5}) is analysed and constraints on the free parameters are obtained by using different sources of data. The viability of such models is analysed by combining the cosmological and local constraints.

%%%%%%%%%%%%%%%%%%%%%%%%%%%%%%%%%%%%%%%%%%%%%%%%%%%%%%%%%%%
\section{FLRW cosmologies in hybrid gravity}
\label{sectionIV}
%%%%%%%%%%%%%%%%%%%%%%%%%%%%%%%%%%%%%%%%%%%%%%%%%%%%%%%%%%%

Let us now study the cosmological evolution of the above model. Here we assume a flat Friedmann-Lema\^itre-Robertson-Walker (FLRW) metric:
\be
ds^2=-dt^2+a(t)^2\sum_{i=1}^{3}dx^{i\ 2}\,,
\label{3.1}
\ee
where $a(t)$ is the scale factor. In this work, we consider a perfect fluid, $T_{\mu\nu}\equiv {\rm diag}(-\rho_m,p_m,p_m,p_m)$, where $\rho_m$ and $p_m$ are the energy density and pressure for the matter content of the universe. Hence, by the field equations (\ref{1.13a}) and (\ref{1.13}), the FLRW equations become \cite{Capozziello:2015lza,Capozziello:2012ny}:
\bea
3H^2=\f{1}{1+\phi}\l[\kappa^2\rho_m+\f{V}{2}-3\dot{\phi}\l(H+\f{\dot{\phi}}{4\phi}\r)\r] \,, \label{3.2a} \\
2\dot{H}=\f{1}{1+\phi}\l[-\kappa^2(\rho_m+p_m)+H\dot{\phi}+\f{3}{2}\f{\dot{\phi}^2}{\phi}-\ddot{\phi}\r] \,,
\label{3.2}
\eea
where $H=\dot{a}/a$ is the Hubble parameter, while the scalar field equation yields:
\be
\ddot{\phi}+3H\dot{\phi}-\f{\dot{\phi}^2}{2\phi}+\f{\phi}{3}\l[2V-(1+\phi) \f{dV}{d\phi}\r]=-\kappa^2\frac{\phi}{3}(\rho_m-3p_m).
\label{3.3}
\ee

In addition, the continuity equation for the matter perfect fluid is given by
\be
\dot{\rho}_m+3H(\rho_m+p_m)=0\ .
\label{3.4}
\ee
Here we are focusing on the late-time epochs, so that we can neglect the radiation contribution and assume a pressureless fluid for the barionic and dark matter content $w_m=p_m/\rho_m=0$. Then, the continuity equation (\ref{3.4}) can be easily solved in terms of the scale factor and the redshift, $z$:
\be
\rho_{m}=\rho_{m0}\l(\frac{a_0}{a}\r)^3=\rho_{m0}\l(1+z\r)^3\ ,
\label{3.5}
\ee
where $1+z=(a_0/a)$ is used. Since our aim is to constrain the model (\ref{2.5}) by using specific observational data, the FLRW equations should be expressed in terms of an observable as the redshift instead of the cosmic time, which leads to:
\bea
H^2 (1+\phi)&=&\Omega_{m}H_0^2 (1+z)^3+\frac{V}{6}
  \nonumber \\  
 &+& \phi' H(1+z)\l(H-(1+z)H\frac{\phi'}{4\phi}\r),
\label{3.6a}
 \eea
 and
 \bea
2HH'(1+z)(1+\phi)=3\Omega_{m}H_0^2 (1+z)^3
   \nn
+2H^2(1+z)\phi'-\frac{3}{2}(1+z)^2H^2\frac{\phi'^{2}}{\phi}
   \nn
+(1+z)^2HH' \phi'+(1+z)^2H^2\phi''\ ,
\label{3.6}
\eea
where $\Omega_m=\kappa^2\rho_{m0}/(3 H_0^2)$ and $H_0$ is the Hubble parameter evaluated today. Hence, we can solve the equations (\ref{3.6a}) and (\ref{3.6}) to draw the cosmological evolution provided by the model (\ref{2.6}). To do so we identify the free parameters of the model as $\{V_0, V_1, \phi_0, \phi_1, H_0\}$, where $\phi_0$ and $\phi_1$ are the scalar field and its first derivative at $z=0$ respectively. However, by the sources of data used in this analysis, the Hubble parameter is dropped out while $V_0$ can also be removed by using the equation (\ref{3.6a}) as a constraint equation to satisfy the flatness condition evaluated at $z=0$:
\be
1=\Omega_m+\Omega_{\phi}\ .
\label{3.7}
\ee
Here $\Omega_{\phi}$ accomplishes for all the extra terms in (\ref{3.6a}). Then, we obtain the following expression relating the free parameters:
\be
V_0=6\l[1+\phi_0-\Omega_m-\phi_1\l(1-\frac{\phi_1}{4\phi_0}\r)\r]-V_1\phi_0^2\ .
\label{3.8}
\ee 
Hence, the only free parameters that remain are $\{\Omega_m, V_1, \phi_0, \phi_1\}$, which are constrained by using Supernovae Ia and Baryonic Acoustic Oscillations (BAO) and the results compared with the local conditions obtained in the section above.

%
%%%%%%%%%%%%%%%%%%%%%%%%%%%%%%%%%%%%%%%%%%%%%%%%%%%%%%
%\section{Observational data}
%%%%%%%%%%%%%%%%%%%%%%%%%%%%%%%%%%%%%%%%%%%%%%%%%%%%%%

In the following, all the observational data used to constrain the model is explained, which are SNe Ia and BAO data. Since the hybrid metric-Palatini gravity model presented here is assumed to represent a late-time parametrization, which recovers $\Lambda$CDM at high redshifts, we do not consider CMB data, particularly the angular scale of the sound horizon and the scale distance to recombination, as they do not provide any extra constraint on the free parameters of the model. In addition, the curvature of the model is assumed to be flat as a prior, since the peaks of the power spectrum from the CMB are very sensible to this value, providing a strict bound given by $\Omega_k<0.005$ \cite{Ade:2015xua}, which can be considered negligible at small redshifts. The analytical form of the $\chi^2$ expression used for each SN and BAO dataset is shown, which will be then minimized to perform the statistical analysis from the Bayesian approach.

%In our paper we don't use CMB data, because integration of the hybrid palatini model upto high redshift will we for sure quite problematic. Knowing that the CMB constrains the curvature to be almost flat, if we consider curvature without the CMB data, the rest of the used data (SN and BAO) may not be enough to properly constrain the curvature. This will imply big errors for the curvature parameter, and if is correlated with other parameters as the referee says, we will also obtain worst results for the other parameters. Besides, fixing the curvature to be flat could also be considered as a "prior" from the unused CMB data, is the first and easiest way to incorporate CMB information to our chains.

%%%%%%%%%%%%%%%%%%%%%%%%%%%%%%%%%%%%%%%%%%%%%%%%%%%%%%%%%%%
\subsection{SNe Ia data}
%%%%%%%%%%%%%%%%%%%%%%%%%%%%%%%%%%%%%%%%%%%%%%%%%%%%%%%%%%%

The Union2.1 compilation \citep{Suzuki:2011hu} dataset was chosen for our SNe Ia test, which comprises of $580$ Type Ia Supernovae distributed in the redshift range $0.015 < z < 1.414$. Besides the distance modulus $\mu(z_i)$ for each SN, the full statistical plus systematics covariance matrix of the survey is also given. The definition of distance modulus is
\be
\mu(z) = 5 \log_{10} d_L(z) + \mu_{0} \;,
\ee
where the dimensionless luminosity distance $d_{L}$ is given by
\be
d_L(z) = (1+z) \int_0^z \frac{dz'}{E(z')} ,
\ee
and $\mu_{0}$ is a nuisance parameter which include all the information related with constants, like the value of Hubble constant $H_{0}$ and the SNeIa absolute magnitude. $E(a)$ is given by the adimensional Hubble function of the model, once it is numerically solved from Eq.~(\ref{3.6a})-(\ref{3.6}). A vector can be defined with the difference between model and observed magnitudes, 
\bea
\textbf{X}_{SN} = \left( \begin{array}{c}
	\mu(z_{1}) - \mu_{obs}(z_{1}) \\
	\ldots \\
	\mu(z_{n}) - \mu_{obs}(z_{n})
\end{array} \right) \; ,
\eea
and we build the SNe contribution to the $\chi^2$ using the covariance matrix $\bf{C}$ from \citep{Suzuki:2011hu},
\be
\chi^2_{SN}=\textbf{X}_{SN}^T \cdot \textbf{C}^{-1} \cdot \textbf{X}_{SN} .
\ee
However, this contribution to the $\chi^2$ from the SNe data contains the nuisance parameter $\mu_{0}$. An analytical marginalization can be done over the $\mu_{0}$ nuisance parameter \citep{Conley:2011ku}, changing the $\chi^2$ contribution of the SNe Ia data, after marginalizing, to the form of
\be
\chi^2_{SN} = a + \log \frac{d}{2 \pi } - \frac{b^2}{d} \, ,
\ee
where 
\bea
a &\equiv&  \textbf{X}_{SN}^{T} \cdot \textbf{C}^{-1} \cdot  \textbf{X}_{SN} \nonumber \\ 
b &\equiv&  \textbf{X}_{SN}^{T} \cdot {\bf C} ^{-1} \cdot \bf{1} \nonumber \\
d &\equiv& {\bf 1}^{T} \cdot {\bf C}^{-1} \cdot \bf{1} \nonumber \, ,
\eea
with $\bf{1}$ standing for the identity matrix and ${ }^{T}$ for transpose.

%%%%%%%%%%%%%%%%%%%%%%%%%%%%%%%%%%%%%%%%%%%%%%%%%%%%%%%%%%%
\subsection{BAO data}
%%%%%%%%%%%%%%%%%%%%%%%%%%%%%%%%%%%%%%%%%%%%%%%%%%%%%%%%%%%

From WiggleZ Dark Energy Survey data \citep{Blake:2012pj}, we have used the Alcock-Paczynski distortion parameter $F(z)$ for the BAO test. This observable is defined as 
\bea
F (z) & \equiv & (1+z) D_A(z) H(z)/c \, ,
\eea
where the angular-diameter distance is 
\be 
D_A(z) = \frac{c}{(1+z)} \int^z_0 \frac{dz'}{H(z')} \, .
\ee 
The defined BAO observables are measured in three different overlapping redshift slices in the WiggleZ survey, where the effective redshift in each bin are $(z_1, \, z_2, \, z_3) = (0.44, \, 0.60, \, 0.73)$. The values that observables take in each redshift bin are 
\bea
F_1 (z_1) &=& 0.482 \nonumber \\
F_2 (z_2) &=& 0.650 \nonumber \\
F_3 (z_3) &=& 0.865 \nonumber \, ,
\eea
and the covariance matrix $10^{3} \, \textbf{C}_{BAO}$ in this case is
\bea
\left( \begin{array}{ccc}
 2.401 & 1.350 & 0.0 \\
 1.350 & 2.809 & 1.934 \\
 0.0   & 1.934 & 5.329 \\
\end{array} \right).
\eea
Thus, the last step to compute the $\chi^2$ contribution of the BAO data is
\be 
\chi^2_{BAO} = (\textbf{X}_{obs} - \textbf{X}_{mod} )^T \textbf{C}_{BAO}^{-1} (\textbf{X}_{obs} - \textbf{X}_{mod} ),
\ee
where $\textbf{X}_{obs} = \left( F_1(z_1),F_2(z_2),F_3(z_3) \right)$ and $\textbf{X}_{mod}$ is the data vector created using the model which is being tested.

\begin{table*}[ht]  %%%%%%%%% Table of EOS parametrizations %%%%%%%%%%%
\begin{center}
\begin{minipage}{0.88\textwidth}
\caption{Constraints for the Palatini Hybrid model together with the $\Lambda$CDM model. Bold parameters are used in MCMC analysis, plain ones are derived. \label{table1}}
\end{minipage}\\
\begin{tabular}{cccccccc}
\hline
\hline
\bf{Model} &  ${\bf V_1/H_0^2}$ & {\boldmath $\phi_0$ } & {\boldmath $\phi_1$}  & $\bf{\Omega_{0m}}$  &  $V_0/H_0^2$ &  $\bf{\chi^2}$ &   $\bf{\chi_{red}^2}$\\
\hline \vspace{-5pt}\\
$\Lambda$CDM  & - & - & - & $0.289_{-0.037}^{+0.041}$ & - & 552.523 & $0.951 $\\
\\
 Hybrid grav. & $-15^{+27}_{-10}           $ & $-0.12\pm 0.62             $& $0.02\pm 0.91$ &  $0.43^{+0.20}_{-0.26} $ & $3.87^{+0.68}_{-0.57}$  & 552.126 & 0.955 \\\\
 \hline \hline
\end{tabular}
\end{center}
\end{table*}

\begin{figure*}[ht]
	\begin{minipage}{1.0\textwidth}
		\centering
		\includegraphics[width=1.0\textwidth]{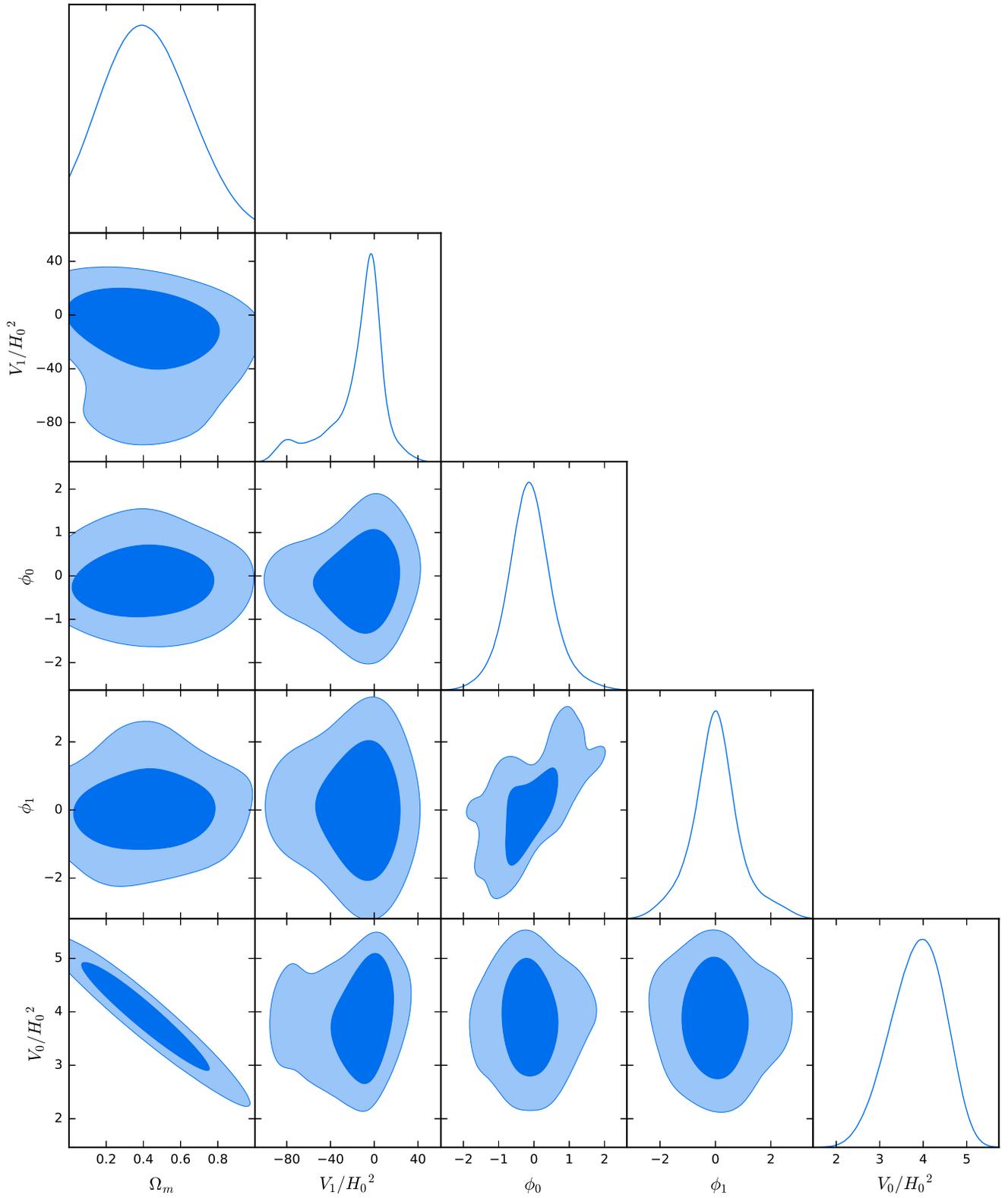}
		\caption{Contours for $1 \sigma$ and $2 \sigma$ with also posterior probabilities of the Palatini Hybrid model.} \label{fig1}
	\end{minipage}
\end{figure*}

\begin{figure*}[ht]
	\begin{minipage}{1.0\textwidth}
		\centering
		\includegraphics[width=0.4\textwidth]{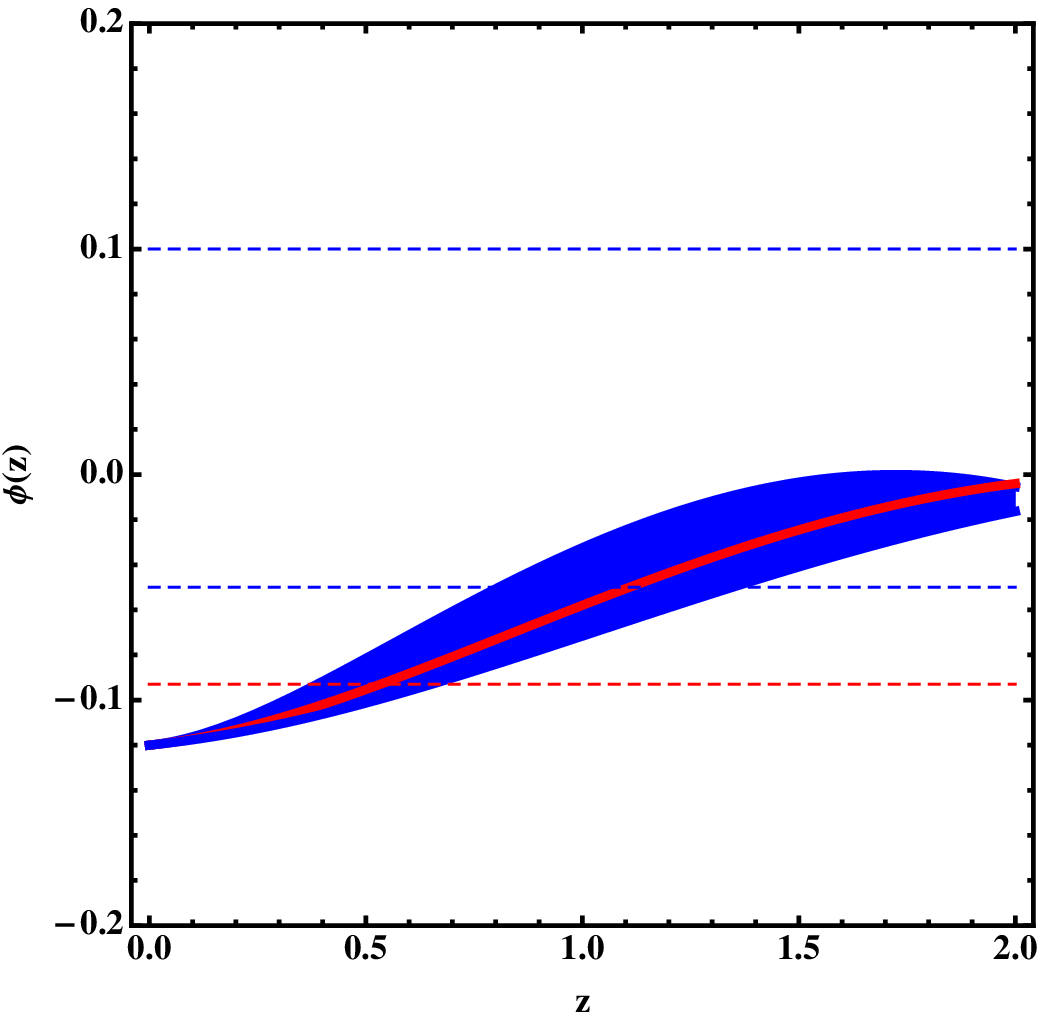}
		\hspace{1cm}		
		\includegraphics[width=0.4\textwidth]{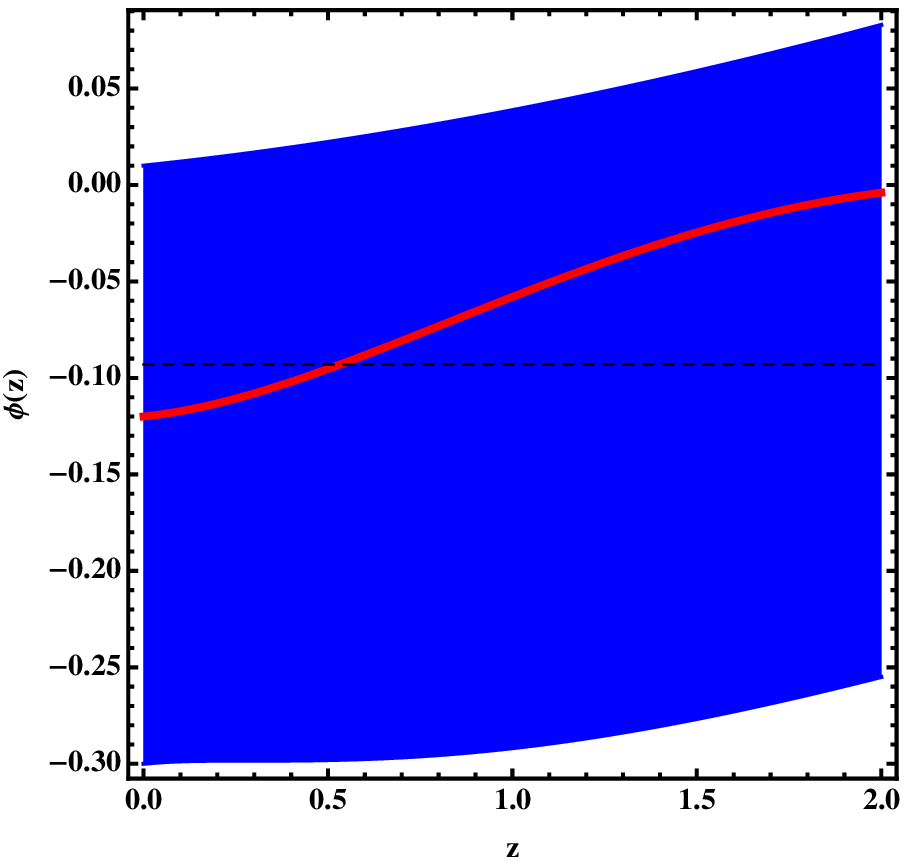}
		\caption{Evolution of the scalar field $\phi=\phi(z)$. Left panel: the blue region represents the $1\sigma$ region for the values of the potential $V_1/H_0^2=-15^{+27}_{-10}$ from table \ref{table1}, the inner red line refers to the mean value. The dashed lines represent the constraint $\phi<V_0/2V_1$ from (\ref{2.7}) for  the mean value (red) and the upper and down limits (blue). Right panel: evolution of the scalar field for different initial conditions $\phi_0$, the red line gives the mean value in table \ref{table1} and the dashed line provides the newtonian constraint \ref{2.7}.} \label{fig2}
	\end{minipage}
\end{figure*}

%%%%%%%%%%%%%%%%%%%%%%%%%%%%%%%%%%%%%%%%%%%%%%%%%%%%%%%%%%%
\section{Results and Discussion}
\label{sectionV}
%%%%%%%%%%%%%%%%%%%%%%%%%%%%%%%%%%%%%%%%%%%%%%%%%%%%%%%%%%%

To perform the statistical Bayesian analysis, the minimisation of the $\chi^2$ is done by using the Markov Chain Monte Carlo (MCMC) Method \citep{Christensen:2001gj,Lewis:2002ah,Trotta:2004qj}. The model contains four free parameters $\{\Omega_m,V_1, \phi_0,\phi_1\}$, where $\phi(z=0)=\phi_0$ and $\phi'(z=0)=\phi_1$ are the initial conditions for the scalar field. The priors used in the MCMC computation are all broad constraints: for the matter we set $0 < \Omega_m < 1$, and for the initial condition parameters $ -10 < \phi_0 < 10 $ and $ -10 < \phi_1 < 10$, while the quadratic term of the potential $V_1/H_0^2$ is left free. The parameter $V_0$ is fixed by the constraint equation (\ref{3.8}). In addition, note that the equations (\ref{3.2a}) and (\ref{3.2}) are not well defined at $\phi=0$, since the equivalence among  the hybrid model (\ref{1.1}) and the scalar-tensor representation (\ref{1.12}) is not longer valid at $\phi=0$, where the conformal factor that relates $R$ and $\mathcal{R}$ diverges. At such point, the theory (\ref{1.1}) reduces to GR with a cosmological constant as can be seen  by the action (\ref{1.7}), i.e., the theory reduces to $\Lambda$CDM in that case, a limit that may be achieved in the model dealt in this paper (\ref{2.5}) when $\mathcal{R}\rightarrow 0$.

The convergence of the MCMC's results were tested using the method of \cite{Dunkley:2004sv}, where almost every parameter has correctly converged. The single parameter that did not fully converge was $V_1/H_0^2$, being its convergence at the edge of acceptable. Nevertheless, all chains have shown the same minimum in the parameter space, even for the $V_1/H_0^2$ parameter. Thus, these results of the MCMC's are summarised in Table \ref{table1}. As shown, the minimum $\chi^2$ achieved by the model is $552.126$, being the reduced $\chi^2$ $0.955$. In comparison to $\Lambda$CDM model, by contrast, when using the same datasets, this results in $\chi^2_{min} = 552.523$, or taking into account the degree of freedom of the model, $\chi^2_{red}=0.951$, with $\Omega = 0.289_{-0.037}^{+0.041}$. Since the $\Lambda$CDM model contains just one free parameter, the reduced $\chi^2_{red}$ becomes smaller. However, the differences on the value of $\chi^2_{red}$ among the models is not significant enough to rule out this model of hybrid gravity. In addition, as pointed above, the hybrid model (\ref{2.5}) contains $\Lambda$CDM as a possible limit. As shown in Table \ref{table1}  and Fig.~\ref{fig2}, the case $\phi_0=0$ is within the $1\sigma$ region, as expected. Moreover, the range values for $\Omega_m$ becomes very large as the auxiliary scalar field may behave as a pressureless fluid for large redshifts.

On the other hand, combining the results of the fits summarised in Table \ref{table1} and the local constraint given by the conditions (\ref{2.7}) to avoid Newtonian corrections, we may conclude whether hybrid gravity may be kept as a viable candidate for dark energy. Note that the initial value for the scalar field should be $\phi_0\ll 1$, such that we restrict the analysis to those values of the $1\sigma$ region in Fig.~\ref{fig1} which accomplishes such constraint. Then, in Fig.~\ref{fig2} the evolution of the scalar field is depicted. In the right panel, the initial condition for $\phi_0$ is fixed while the scalar potential varies according to the values of the $1\sigma$ region for $V_1$ and $V_0$, while the dashed lines represent the constraint (\ref{2.7}) on the mass of the scalar field, which varies according to $V_1$ and $V_0$. Here we have assumed the mean value in Table \ref{table1} for $\phi_0$. In the right panel, the scalar potential is fixed at $V_1=-15$ and the initial value of the scalar field varies. As shown in the right panel, the scalar field tends to zero at large redshifts, converging to the $\Lambda$CDM model in the past, but only some values of the scalar potential satisfy the constraint (\ref{2.7}). Hence, whether the quadratic term of the potential $V_1<0$, the initial value of the scalar field should be $\phi_0>|V_0/2V_1|$, and close to zero, while the initial conditions can be set more freely when $V_1>0$. \\

Nevertheless, even small values of the scalar field $\phi$, together with a small mass, may induce corrections on the gravitational constant and the post-Newtonian parameter (\ref{2.4}). According to recent experiments, the gravitational constant is given by $G=6.674 08\times 10^{-11} \,  m^3 kg^{-1} s^{-2} $ with an uncertainty of approximately $0.05\%$, which means $500$ parts per million \cite{Mohr:2015ccw}. On the other hand, the Cassini tracking estimates a value of $\gamma=1$ with a very small deviation \cite{Bertotti:2003rm,Will:2005va}. Hence, the following constraints should also hold in order to satisfy the experiments requirements:
\be
\left|\frac{G_{eff}-G}{G}\right|<4.7\times 10^{-5}\ , \quad \left|\gamma-1\right|<2.3\times 10^{-5}\ .
\label{Results1}
\ee
Hence, the deviations in our model, which are expressed by Eqs. (\ref{2.4}) will impose a small enough value of the scalar field to satisfy the constraint on $G$, whereas for the post-Newtonian parameter $\gamma$, a sufficiently large mass of the scalar field may be enough. For the model studied in this paper, the scalar field mass is given by Eq. (\ref{2.7}), which in units of length can be rewritten as follows:
\be
m_{\phi}^2=\frac{2}{3}\left(v_0-2v_1\phi\right)H_0^2\ .
\label{Results2}
\ee
Here, $v_0=V_0/H_0^2$ and $v_1=V_1/H_0^2$. Let us first consider two particular extreme cases:
\begin{itemize}
\item For a large mass, $m_{\phi}r \gg 1$, so $e^{-m_{\phi}r}\approx 0$, it yields:
\bea
\left|\frac{G_{eff}-G}{G}\right|&\approx& \left|\frac{1}{1+\phi_0}-1\right|<4.7\times 10^{-5}\ , \nn
 \left|\gamma-1\right|&=& \left|\frac{2\phi_0e^{-m_{\phi}r}/3}{1-(\phi_0 e^{-m_{\phi}r}/3)}\right| \approx 0\ .
\label{Results5}
\eea
Hence, the constraint on the post-Newtonian parameter is satisfied as far as the mass of the scalar field is large (and the value of the scalar field itself small), so the constraint on the gravitational constant provides the bounds on the scalar field:
\be
-5\times 10^{-4}<\phi_0<5\times 10^{-4}\ .
\label{Results6}
\ee
\item For a small mass, $m_{\phi}r \ll 1$, so $e^{-m_{\phi}r}\approx 1$. Then, the corrections yield:
\bea
\left|\frac{G_{eff}-G}{G}\right|&\approx& \left|\left(1-\frac{\phi_0}{3}\right)\frac{1}{1+\phi_0}-1\right|<4.7\times 10^{-5}\ , \nn
 \left|\gamma-1\right|&\approx& \left|\frac{2\phi_0/3}{1-(\phi_0/3)}\right| <2.3\times 10^{-5}\ .
\label{Results3}
\eea
The second equation is much stricter than the first, so the following bounds on the scalar field are obtained:
\be
-3.4\times 10^{-5}<\phi_0<3.4\times 10^{-5} \,.
\label{Results4}
\ee
\
\end{itemize}

We can now combine the above restrictions with the constraints obtained by using cosmology. By analysing the 1$\sigma$ region of the free parameters of the model depicted in Fig.~\ref{fig1} and summarised in Table \ref{table1}, the mass of the scalar field turns out to be $m_{\phi}=1-3H_0$ at the present time, which leads to a very small mass and $m_{\phi}r \ll 1$ at local scales as the Earth or the Solar System. Hence, in order to avoid corrections on the gravitational constant, neither on $\gamma$, the constraint (\ref{Results4}) should be satisfied. In this sense, Fig.~\ref{fig1} shows that the value of the scalar field today $\phi_0$ can be set as small as required, so the above constraint (\ref{Results4}) is satisfied. In addition, Fig.~\ref{fig2} shows that for the range of redshifts studied in the paper (and where the model is assumed to be effective), the evolution of the scalar field can be kept small to accomplish the above requirement. Nevertheless, local constraints impose much stricter conditions on our model.

%
%\begin{table}[h]
%	\caption{Constraints for the Palatini Hybrid model. Bold parameters are used in MCMC, plain ones are derived.
%	}\label{table-UDM}
%	\begin{tabular} { l  c}
%		\hline
%		Parameter &  68\% limits\\
%		\hline
%		{\boldmath$\Omega_m       $} & $0.43^{+0.20}_{-0.26}      $\\
%		
%		{\boldmath$V_1/H_0        $} & $-15^{+27}_{-10}           $\\
%		
%		{\boldmath$\phi_0         $} & $-0.12\pm 0.62             $\\
%		
%		{\boldmath$\phi_1         $} & $0.02\pm 0.91              $\\
%		
%		$V_0/H_0                   $ & $3.87^{+0.68}_{-0.57}      $\\
%		\hline
%	\end{tabular}
%\end{table}

%\vspace{-2pt}

%%%%%%%%%%%%%%%%%%%%%%%%%%%%%%%%%%%%%%%%%%%%%%%%%%%%%%%%%%%
\section{Conclusions}\label{sectionConclusion}
%%%%%%%%%%%%%%%%%%%%%%%%%%%%%%%%%%%%%%%%%%%%%%%%%%%%%%%%%%%

In this paper we have focused on the analysis of the so-called hybrid gravity, where GR is slightly modified by including a non-metric curvature term. Then, by using an auxiliary scalar field, one can easily show that such a theory of gravity is equivalent to a Brans-Dicke-like action, such that the analysis turns out easier. Nevertheless, such an equivalence breaks down in the limit where GR is recovered. By studying the weak field limit of the theory, the effective Newton constant and the post-Newtonian parameters are obtained, which depend not only on the mass of the scalar field (as occurs in chameleon fields) but on the background value of the scalar field itself, as shown in Ref.~\cite{Harko:2011nh}, such that a small enough value of the scalar field may be enough to avoid violations at local scales, in contrast to the large mass required in the chameleon mechanism. Then, a simple model is proposed where the constraints on the free parameters of the theory are obtained.

The cosmological evolution of the model is studied by assuming a flat FLRW universe, where the auxiliary scalar field is responsible for the late-time cosmic acceleration, or in other words, the non-metric part of the action. By using Supernovae Ia data and Baryonic Acoustic Oscillations data, the free parameters of the model are fitted through an MCMC analysis. The values of $\chi^2$ and $\chi^2_{red}$ are obtained and compared to the $\Lambda$CDM model. The values of $\chi_{red}^2$ do not differ significantly, such that hybrid gravity remains as a possible candidate for dark energy. Moreover, the quadratic term of the scalar potential is well constrained, concluding that it would likely be greater than zero. Nevertheless, the errors on the free parameters turn out much larger, particularly on the value of the matter density $\Omega_m$, since the number of parameters is larger than in the $\Lambda$CDM model and the scalar field may mimic a pressureless fluid at large redshifts. 

However, when comparing the cosmological constraints to the local ones, we found out that the value of the scalar field today determines whether the cosmological constraints and the local ones are both satisfied at the same time, since the mass of the scalar field remains very small within the 1$\sigma$ region of the free parameters. Hence, the value of the scalar field today $\phi_0$ becomes much better constrained, reducing its viable range noticeably around zero, where General Relativity with a cosmological constant is recovered. In this sense, the model turns out to be suitable for passing the local and cosmological restrictions satisfactorily. 

Hence, we can conclude that modifications of GR in the way of breaking the metricity condition are not ruled out, but severe constraints are imposed. Future analysis of the perturbations and the growth of large scale structure may provide additional information and constraints on this type of models.

%%%%%%%%%%%%%%%%%%%%%%%%%%%%%%%%%%%%%%%%%%%%%%%%%%%%%%%%%%%
\section*{Acknowledgments}

FSNL acknowledges financial support of the Funda\c{c}\~{a}o para a Ci\^{e}ncia e Tecnologia through an Investigador FCT Research contract, with reference IF/00859/2012, funded by FCT/MCTES (Portugal).
DSG is funded by the Juan de la Cierva program (Spain) No.~IJCI-2014-21733 and by MINECO (Spain), project FIS2013-44881. IL acknowledges financial support from the University of the Basque Country UPV/EHU PhD grant 750/2014. IL also thanks the hospitality of the Instituto de Astrof\'{\i}sica e Ci\^{e}ncias do Espa\c{c}o, Faculdade de Ci\^encias da Universidade de Lisboa where part of this work was carried out.
This article is based upon work from COST Action CA15117, supported by COST (European Cooperation in Science and Technology).
%%%%%%%%%%%%%%%%%%%%%%%%%%%%%%%%%%%%%%%%%%%%%%%%%%%%%%%%%%%

%%%%%%%%%%%%%%%%%%%%%%%%%%%%%%%%%%%%%%%%%%%%%%%%%%%%%%%%%%%


\begin{thebibliography}{999}
%%%%%%%%%%%%%%%%%%%%%%%%%%%%%%%%%%%%%%%%%%%%%%%%%%%%%%%%%%%

%\cite{Perlmutter:1998np}
\bibitem{Perlmutter:1998np} 
  S.~Perlmutter {\it et al.} [Supernova Cosmology Project Collaboration],
  %``Measurements of Omega and Lambda from 42 high redshift supernovae,''
  Astrophys.\ J.\  {\bf 517}, 565 (1999)
  [astro-ph/9812133].

%\cite{Riess:1998cb}
\bibitem{Riess:1998cb} 
  A.~G.~Riess {\it et al.} [Supernova Search Team Collaboration],
  %``Observational evidence from supernovae for an accelerating universe 
  %  and a cosmological constant,''
  Astron.\ J.\  {\bf 116}, 1009 (1998)
  [astro-ph/9805201].
  %%CITATION = doi:10.1086/300499;%%

%\cite{Copeland:2006wr}
\bibitem{Copeland:2006wr} 
  E.~J.~Copeland, M.~Sami, and S.~Tsujikawa,
  %``Dynamics of dark energy,''
  Int.\ J.\ Mod.\ Phys.\ D {\bf 15}, 1753 (2006)
  [hep-th/0603057].

%\cite{Sotiriou:2008rp}
\bibitem{Sotiriou:2008rp} 
  T.~P.~Sotiriou and V.~Faraoni,
  %``$f(R)$ Theories of gravity,''
  Rev.\ Mod.\ Phys.\  {\bf 82}, 451 (2010)
  %doi:10.1103/RevModPhys.82.451
  [arXiv:0805.1726 [gr-qc]].
  
%\cite{DeFelice:2010aj}
\bibitem{DeFelice:2010aj} 
  A.~De Felice and S.~Tsujikawa,
 % ``$f(R)$ theories,''
  Living Rev.\ Rel.\  {\bf 13}, 3 (2010)
  [arXiv:1002.4928 [gr-qc]].  

%\cite{Clifton:2011jh}
\bibitem{Clifton:2011jh} 
  T.~Clifton, P.~G.~Ferreira, A.~Padilla, and C.~Skordis,
  %``Modified gravity and cosmology,''
  Phys.\ Rept.\  {\bf 513}, 1 (2012)
  [arXiv:1106.2476 [astro-ph.CO]].
  
\bibitem{Nojiri:2010wj} 
  S.~Nojiri and S.~D.~Odintsov,
  %``Unified cosmic history in modified gravity: from 
   %$F(R)$ theory to Lorentz non-invariant models,''
  Phys.\ Rept.\  {\bf 505}, 59 (2011)
  [arXiv:1011.0544 [gr-qc]].

\bibitem{bookFR}S.~Capozziello and V.~Faraoni, \textit{Beyond Einstein Gravity} (Springer, Dordrecht, 2010).


\bibitem{reviewAandD}A.~de la Cruz-Dombriz and D.~S\'aez-G\'omez, Entropy {\bf 14}, 1717 (2012);


\bibitem{Olmo:2011uz} 
  G.~J.~Olmo,
  %``Palatini Approach to Modified Gravity: f(R) Theories and Beyond,''
  Int.\ J.\ Mod.\ Phys.\ D {\bf 20}, 413 (2011)
  %doi:10.1142/S0218271811018925
  [arXiv:1101.3864 [gr-qc]].
%ÒPalatini f(R) black holes in nonlinear electrodynamicsÓ;
G. J. Olmo and D.~Rubiera-Garcia,
Phys. Rev. D 84 (2011) 124059;
%Wormholes and nonsingular space-times in Palatini f(R) gravity;
C. Bambi, A. Cardenas-Avendano, G. J. Olmo, and D.~Rubiera-Garcia,
Phys. Rev. D 93 (2016) no.6, 064016;
%ÒNonsingular black holes in f(R) theoriesÓ;
G. J. Olmo and D.~Rubiera-Garcia,
Universe 2015, no.2, 173-185;
C.~Bejarano, G.~J.~Olmo and D.~Rubiera-Garcia,
  %``What is a singular black hole beyond General Relativity?,''
  Phys.\ Rev.\ D {\bf 95}, no. 6, 064043 (2017)
  doi:10.1103/PhysRevD.95.064043
  [arXiv:1702.01292 [hep-th]].


\bibitem{Hu:2007nk} 
  W.~Hu and I.~Sawicki,
  %``Models of f(R) Cosmic Acceleration that Evade Solar-System Tests,''
  Phys.\ Rev.\ D {\bf 76}, 064004 (2007)
  %doi:10.1103/PhysRevD.76.064004
  [arXiv:0705.1158 [astro-ph]];
%  \bibitem{Nojiri:2007as} 
  S.~Nojiri and S.~D.~Odintsov,
  %``Unifying inflation with LambdaCDM epoch in modified f(R) gravity consistent with Solar System tests,''
  Phys.\ Lett.\ B {\bf 657}, 238 (2007)
 % doi:10.1016/j.physletb.2007.10.027
  [arXiv:0707.1941 [hep-th]].

\bibitem{delaCruz-Dombriz:2015tye} 
  A.~de la Cruz-Dombriz, P.~K.~S.~Dunsby, S.~Kandhai and D.~Saez-Gomez,
  %``Theoretical and observational constraints of viable f(R) theories of gravity,''
  Phys.\ Rev.\ D {\bf 93}, no. 8, 084016 (2016)
  %doi:10.1103/PhysRevD.93.084016
  [arXiv:1511.00102 [gr-qc]].


%\cite{Harko:2011nh}
\bibitem{Harko:2011nh} 
  T.~Harko, T.~S.~Koivisto, F.~S.~N.~Lobo and G.~J.~Olmo,
  %``Metric-Palatini gravity unifying local constraints and late-time cosmic acceleration,''
  Phys.\ Rev.\ D {\bf 85}, 084016 (2012)
  %doi:10.1103/PhysRevD.85.084016
  [arXiv:1110.1049 [gr-qc]].
  %%CITATION = doi:10.1103/PhysRevD.85.084016;%%
  %56 citations counted in INSPIRE as of 27 Apr 2016
  
\bibitem{Capozziello:2012ny} 
  S.~Capozziello, T.~Harko, T.~S.~Koivisto, F.~S.~N.~Lobo and G.~J.~Olmo,
  %``Cosmology of hybrid metric-Palatini f(X)-gravity,''
  JCAP {\bf 1304}, 011 (2013)
  %doi:10.1088/1475-7516/2013/04/011
  [arXiv:1209.2895 [gr-qc]].
  
  %\cite{Carloni:2015bua}
\bibitem{Carloni:2015bua} 
  S.~Carloni, T.~Koivisto and F.~S.~N.~Lobo,
  %``Dynamical system analysis of hybrid metric-Palatini cosmologies,''
  Phys.\ Rev.\ D {\bf 92}, no. 6, 064035 (2015)
  %doi:10.1103/PhysRevD.92.064035
  [arXiv:1507.04306 [gr-qc]].

  \bibitem{Capozziello:2012qt} 
  S.~Capozziello, T.~Harko, T.~S.~Koivisto, F.~S.~N.~Lobo and G.~J.~Olmo,
  %``The virial theorem and the dark matter problem in hybrid metric-Palatini gravity,''
  JCAP {\bf 1307}, 024 (2013)
  %doi:10.1088/1475-7516/2013/07/024
  [arXiv:1212.5817 [physics.gen-ph]].
  
  \bibitem{Capozziello:2013yha} 
  S.~Capozziello, T.~Harko, T.~S.~Koivisto, F.~S.~N.~Lobo and G.~J.~Olmo,
  %``Galactic rotation curves in hybrid metric-Palatini gravity,''
  Astropart.\ Phys.\  {\bf 50-52}, 65 (2013)
  %doi:10.1016/j.astropartphys.2013.09.005
  [arXiv:1307.0752 [gr-qc]].

  \bibitem{Capozziello:2013uya} 
  S.~Capozziello, T.~Harko, F.~S.~N.~Lobo and G.~J.~Olmo,
  %``Hybrid modified gravity unifying local tests, galactic dynamics and late-time cosmic acceleration,''
  Int.\ J.\ Mod.\ Phys.\ D {\bf 22}, 1342006 (2013)
  %doi:10.1142/S0218271813420066
  [arXiv:1305.3756 [gr-qc]].

\bibitem{Capozziello:2015lza} 
  S.~Capozziello, T.~Harko, T.~S.~Koivisto, F.~S.~N.~Lobo and G.~J.~Olmo,
  %``Hybrid metric-Palatini gravity,''
  Universe {\bf 1}, no. 2, 199 (2015)
  %doi:10.3390/universe1020199
  [arXiv:1508.04641 [gr-qc]].

%\cite{Lima:2014aza}
\bibitem{Lima:2014aza} 
  N.~A.~Lima,
  %``Dynamics of Linear Perturbations in the hybrid metric-Palatini gravity,''
  Phys.\ Rev.\ D {\bf 89}, no. 8, 083527 (2014)
  %doi:10.1103/PhysRevD.89.083527
  [arXiv:1402.4458 [astro-ph.CO]].
  
  %\cite{Lima:2015nma}
\bibitem{Lima:2015nma} 
  N.~A.~Lima and V.~S.-Barreto,
  %``Constraints on Hybrid Metric-palatini Gravity from Background Evolution,''
  Astrophys.\ J.\  {\bf 818}, no. 2, 186 (2016)
  %doi:10.3847/0004-637X/818/2/186
  [arXiv:1501.05786 [astro-ph.CO]].


%Palatini f(R) references

 

\bibitem{Olmo:2005hc} 
  G.~J.~Olmo,
  %``Post-Newtonian constraints on f(R) cosmologies in metric and Palatini formalism,''
  Phys.\ Rev.\ D {\bf 72}, 083505 (2005)
  %doi:10.1103/PhysRevD.72.083505
  [gr-qc/0505135];
%\bibitem{Capozziello:2007ms} 
  S.~Capozziello, A.~Stabile and A.~Troisi,
  %``The Newtonian Limit of f(R) gravity,''
  Phys.\ Rev.\ D {\bf 76}, 104019 (2007)
  %doi:10.1103/PhysRevD.76.104019
  [arXiv:0708.0723 [gr-qc]].


  
 \bibitem{Khoury:2003rn} 
  J.~Khoury and A.~Weltman,
  %``Chameleon cosmology,''
  Phys.\ Rev.\ D {\bf 69}, 044026 (2004)
  %doi:10.1103/PhysRevD.69.044026
  [astro-ph/0309411];
%  \bibitem{Khoury:2003aq} 
  J.~Khoury and A.~Weltman,
  %``Chameleon fields: Awaiting surprises for tests of gravity in space,''
  Phys.\ Rev.\ Lett.\  {\bf 93}, 171104 (2004)
  %doi:10.1103/PhysRevLett.93.171104
  [astro-ph/0309300]. 
  

  

  
  %Sources observational data refs
  
  
  \bibitem{Ade:2015xua} 
  P.~A.~R.~Ade {\it et al.} [Planck Collaboration],
  %``Planck 2015 results. XIII. Cosmological parameters,''
  Astron.\ Astrophys.\  {\bf 594}, A13 (2016)
  %doi:10.1051/0004-6361/201525830
  [arXiv:1502.01589 [astro-ph.CO]].
  
  	\bibitem{Suzuki:2011hu} 
	N.~Suzuki {\it et al.},
	%``The Hubble Space Telescope Cluster Supernova Survey: V. Improving the Dark Energy Constraints Above z>1 and Building an Early-Type-Hosted Supernova Sample,''
	Astrophys.\ J.\  {\bf 746}, 85 (2012)
	%doi:10.1088/0004-637X/746/1/85
	[arXiv:1105.3470 [astro-ph.CO]].
	%%CITATION = doi:10.1088/0004-637X/746/1/85;%%
	%768 citations counted in INSPIRE as of 21 Dec 2016
	
	%\cite{Conley:2011ku}
	\bibitem{Conley:2011ku} 
	A.~Conley {\it et al.} [SNLS Collaboration],
	%``Supernova Constraints and Systematic Uncertainties from the First 3 Years of the Supernova Legacy Survey,''
	Astrophys.\ J.\ Suppl.\  {\bf 192}, 1 (2011)
	%doi:10.1088/0067-0049/192/1/1
	[arXiv:1104.1443 [astro-ph.CO]].
	%%CITATION = doi:10.1088/0067-0049/192/1/1;%%
	%381 citations counted in INSPIRE as of 21 Dec 2016
	
	%\cite{Blake:2012pj}
	\bibitem{Blake:2012pj} 
	C.~Blake {\it et al.},
	%``The WiggleZ Dark Energy Survey: Joint measurements of the expansion and growth history at z < 1,''
	Mon.\ Not.\ Roy.\ Astron.\ Soc.\  {\bf 425}, 405 (2012)
	%doi:10.1111/j.1365-2966.2012.21473.x
	[arXiv:1204.3674 [astro-ph.CO]].
	%%CITATION = doi:10.1111/j.1365-2966.2012.21473.x;%%
	%253 citations counted in INSPIRE as of 21 Dec 2016
	

	%\cite{Christensen:2001gj}
	\bibitem{Christensen:2001gj} 
	N.~Christensen, R.~Meyer, L.~Knox and B.~Luey,
	%``II. Bayesian methods for cosmological parameter estimation from cosmic microwave background measurements,''
	Class.\ Quant.\ Grav.\  {\bf 18}, 2677 (2001)
	%doi:10.1088/0264-9381/18/14/306
	[astro-ph/0103134].
	%%CITATION = doi:10.1088/0264-9381/18/14/306;%%
	%99 citations counted in INSPIRE as of 21 Dec 2016
	
	%\cite{Lewis:2002ah}
	\bibitem{Lewis:2002ah} 
	A.~Lewis and S.~Bridle,
	%``Cosmological parameters from CMB and other data: A Monte Carlo approach,''
	Phys.\ Rev.\ D {\bf 66}, 103511 (2002)
	%doi:10.1103/PhysRevD.66.103511
	[astro-ph/0205436].
	%%CITATION = doi:10.1103/PhysRevD.66.103511;%%
	%1889 citations counted in INSPIRE as of 21 Dec 2016
	
	%\cite{Trotta:2004qj}
	\bibitem{Trotta:2004qj} 
	R.~Trotta,
	``Cosmic microwave background anisotropies: Beyond standard parameters,''
	astro-ph/0410115.
	%%CITATION = ASTRO-PH/0410115;%%
	%15 citations counted in INSPIRE as of 21 Dec 2016
  
	%\cite{Dunkley:2004sv}
	\bibitem{Dunkley:2004sv} 
	J.~Dunkley, M.~Bucher, P.~G.~Ferreira, K.~Moodley and C.~Skordis,
	%``Fast and reliable mcmc for cosmological parameter estimation,''
	Mon.\ Not.\ Roy.\ Astron.\ Soc.\  {\bf 356}, 925 (2005)
	%doi:10.1111/j.1365-2966.2004.08464.x
	[astro-ph/0405462].
	%%CITATION = doi:10.1111/j.1365-2966.2004.08464.x;%%
	%160 citations counted in INSPIRE as of 30 Jan 2017
	
	\bibitem{Mohr:2015ccw} 
  P.~J.~Mohr, D.~B.~Newell and B.~N.~Taylor,
  %``CODATA Recommended Values of the Fundamental Physical Constants: 2014,''
  Rev.\ Mod.\ Phys.\  {\bf 88}, no. 3, 035009 (2016)
  %doi:10.1103/RevModPhys.88.035009
  [arXiv:1507.07956 [physics.atom-ph]].
	
	\bibitem{Bertotti:2003rm} 
  B.~Bertotti, L.~Iess and P.~Tortora,
  %``A test of general relativity using radio links with the Cassini spacecraft,''
  Nature {\bf 425}, 374 (2003).
  %doi:10.1038/nature01997
  
  \bibitem{Will:2005va} 
  C.~M.~Will,
  %``The Confrontation between general relativity and experiment,''
  Living Rev.\ Rel.\  {\bf 9}, 3 (2006)
  %doi:10.12942/lrr-2006-3
  [gr-qc/0510072].
	
	
	
\end{thebibliography}
\end{document}